\documentclass[10pt,twocolumn]{article}

\usepackage{graphicx}
\usepackage{amsfonts}
\usepackage{amsmath}
\usepackage{amssymb}
\usepackage{mathrsfs} % cursive letters
\usepackage{lipsum}
\usepackage{color}
\usepackage{siunitx}
\usepackage{authblk}
\usepackage[margin=2cm]{geometry}
\usepackage{mathtools, cuted}

\usepackage{bm}
\usepackage{amsmath}

\title{Convective evaporation of vertical films}
\author[1]{Fran\c{c}ois Boulogne}
\author[2]{Benjamin Dollet}
\affil[1]{Laboratoire de Physique des Solides, CNRS, Univ. Paris-Sud, Universit\'e Paris-Saclay, Orsay 91405, France}
\affil[2]{Univ. Grenoble Alpes, CNRS, LIPhy, 38000 Grenoble, France}

\date{\today}
\begin{document}

\twocolumn[
    \begin{@twocolumnfalse}
        \maketitle
        \begin{abstract}
        %\TODO{\lipsum[1]{}}
            Motivated by the evaporation of soap films, which has a significant effect on their lifetime, we performed an experimental study on the evaporation of vertical surfaces with model systems based on hydrogels.
            From the analogy between heat and mass transfer, we adopt a model describing the natural convection in the gas phase due to a density contrast between dry and saturated air.
            Our measurements show a good agreement with this model, both in terms of scaling law with the Grashof number and in terms of order of magnitude.
            We discuss the corrections to take into account, notably the contribution of edge effects, which have a small but visible contribution when lateral and bottom surface areas are not negligible compared to the main evaporating surface area.
        \end{abstract}
    \end{@twocolumnfalse}
]

%%%%%%%%%%%%%%%%%%%%%%%%%%%%%%
%
% INTRODUCTION
%
%%%%%%%%%%%%%%%%%%%%%%%%%%%%%%
\section{Introduction}

\begin{figure}
    \centering
    \includegraphics[width=\linewidth]{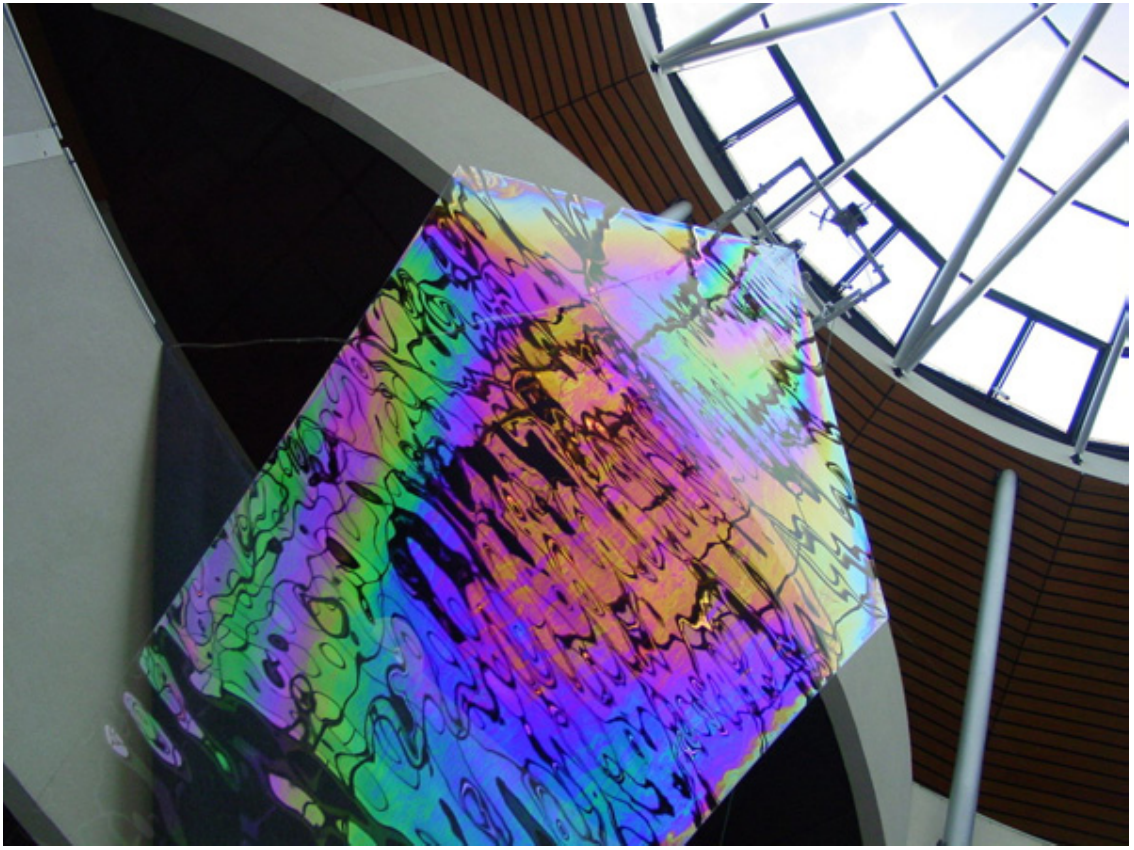}
    \caption{Giant soap film of $10$~m height that brings the question of evaporation on its stability.
The picture is extracted from Ref. \cite{Ballet2006}.
     \copyright European Physical Society. Reproduced by permission of IOP Publishing. All rights reserved. Credit F. Mondot. The photograph was taken at Collège Villeneuve, Grenoble.
    }
    \label{fig:giant_soap_film}
\end{figure}

Giant soap films are spectacular structures since they are uncommon in nature \cite{Rutgers2001,Ballet2006,Cohen2017}.
Indeed, soap films are well-known to be fragile and short standing.
To  maintain such large films artificially, the soap solution is constantly injected at the top to balance the effects of drainage and evaporation \cite{Rutgers2001,Ballet2006}.
These are common in scientific outreach events, as shown in Fig.~\ref{fig:giant_soap_film} where they attract interest from the public due to its gigantic size.
Film evaporation depends on the environmental conditions such as the weather or air currents in the room. Thus, daily fluctuations of these parameters influence the success of the demonstration.

%In scientific outreach events, giant soap films such as shown in Fig.~\ref{fig:giant_soap_film} are known to be spectacular and to attract interest from the public.
%Soap films are well-known to be fragile and short standing.
%To maintain such large films, the soap solution is constantly injected at the top to balance the effects of drainage and evaporation \cite{Rutgers2001,Ballet2006}.
%Indeed, evaporation depends on the environmental conditions such as the weather or air flows in the room.
%Thus, daily fluctuations of these parameters influence the success of the demonstration.

Several scientific studies report effects of evaporation on the soap film stability \cite{Li2010}.
Yaminsky \textit{et al.} evidenced that impurities influence the lifetime of soap film only if placed in an unsaturated atmosphere \cite{Yaminsky2010}.
In addition, Li \textit{et al.} pointed out that the nonuniformity of the evaporative flux induces Marangoni effects, which trigger the film burst \cite{Li2012}.
Moreover, chemistry can have a significant role on the evaporation kinetics as dense monolayers at interfaces can reduce significantly the evaporation rate \cite{Mer1962,Langevin2000}.
Therefore, the complexity in studying evaporation on soap films arises from the interplay of chemistry, hydrodynamics of liquid films with interfacial effects, and the dynamics of evaporation.

In 1887, Maxwell described the evaporation of a suspended spherical drop as a diffusive process of water in the gas phase.
While this description is successful in many situations such as the evaporation of small sessile drops \cite{Deegan1997,Stauber2014}, natural convection has been recently revealed in experimental studies with various liquids \cite{Shahidzadeh-Bonn2006,Dunn2009,Weon2011,Kelly-Zion2011,Kelly-Zion2013,Carle2013a,Somasundaram2015,Carle2016,Boulogne2017a}.

Indeed, water vapor is less dense than dry air.
Above a critical size of the evaporative surface, a convective gas flow takes place, which affects the vapor concentration field.
Typically, the critical size above which convection can be expected is about a few millimeters in room conditions for water.
Models have been developed to describe particular configurations such as the evaporation of pendant drops \cite{Dehaeck2014} and horizontal circular disks \cite{Dollet2017}. Similar convection effects occur in the dissolution of sessile drops \cite{Dietrich2016,Laghezza2016}.

In this paper, we aim to describe the convective evaporation of vertical films.
First, based on an analogy with heat transfer on flat vertical surfaces, we present the theory describing the natural convective flow in such geometry with a prefactor determined numerically.
Because the evaporation of soap films is particularly challenging to study due to their limited lifetime and to the complex interplay between chemical and hydrodynamical effects, we make the choice to perform experiments on a model system, which consists of hydrogels.
We compare our measurements to the theoretical prediction and we concluded on the significance for future soap films studies.

% \paragraph{More to cite?}
% \begin{itemize}
%     \item \cite{Wagner1949} Paper on natural convection in dissolution
%         \item \cite{Rio2014} review Evap ?
%     \item \cite{Corkill1963}
% \end{itemize}

\section{Experimental methods}

\subsection{Hydrogels}

\begin{figure*}
    \centering
    \includegraphics[scale=1]{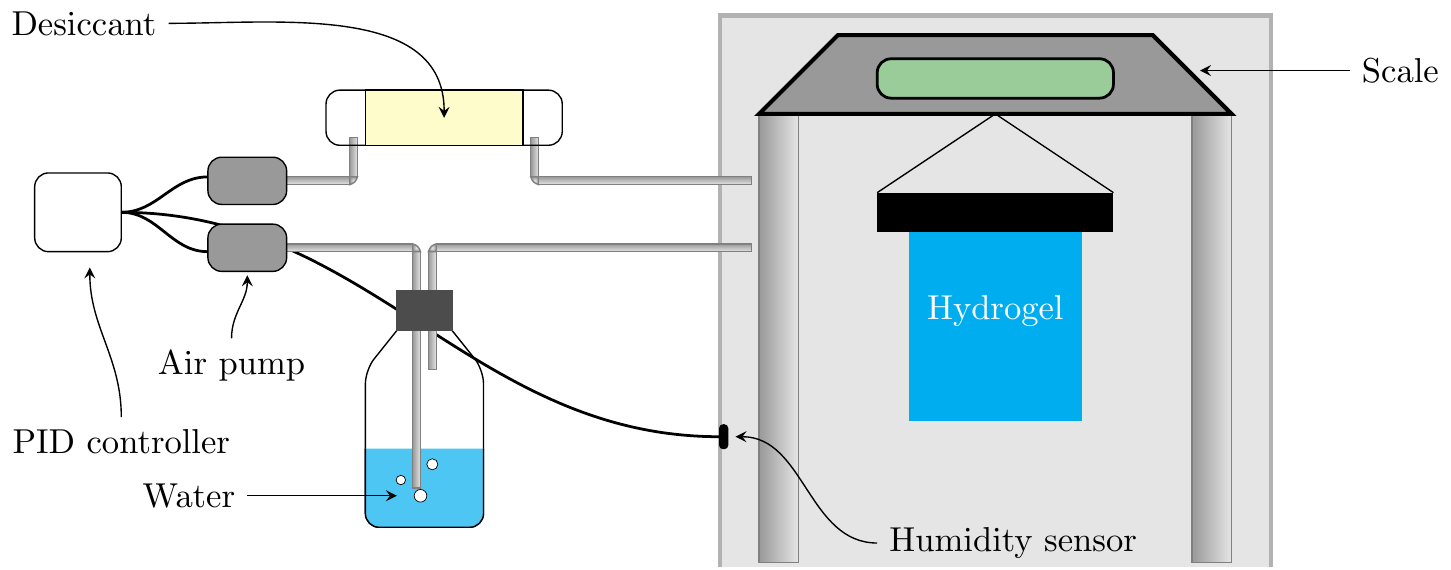}
    \caption{Schematics of the experimental setup.
    The left part shows the principle of the humidity controller and the right part is the box in which a hydrogel is hung on a precision scale mounted on pillars.}
    \label{fig:exp_setup}
\end{figure*}

Hydrogels are synthesized in polydimethyl acrylamid (PDMA) \cite{Sudre2011}.
The quantity of the monomer (N,N-dimethylacrylamide denoted DMA) is set by the mass ratio $m_{\rm DMA} / (m_{\rm DMA} + m_{w}) = 0.1$, where $m_w$ is the mass of pure water.
The crosslinker (N,N'-methylene-bis-acrylamide denoted MBA) is added at a molar concentration ratio of $[\rm{MBA}]/[\rm{DMA}] = 2\times 10^{-2}$.
Finally, a quantity of initiators (potassium persulfate and N,N,N',N'-tetramethylethylenediamine) is set to a molar ratio of 1\% of the monomer quantity. All the chemical are purchased from Sigma-Aldrich (France).

The solution is poured in cells made of two glass slides separated by a rubber of about $3$~mm in thickness, and hold with drawing clips.
The cell is sealed with a piece of parafilm and left in room conditions for 24~h to complete the reaction.
Then, the gel is carefully unmolded and placed in a tank filled with pure water for three days to swell the hydrogel.

The hydrogel is cut with a razor blade in a rectangular shape of the desired size.
The swollen hydrogel has a thickness $b=4.5\pm0.5$~mm.

\subsection{Evaporative flux measurements}

The full setup is represented in figure~\ref{fig:exp_setup}.
Our experiments of controlled evaporation are performed in a box made in polycarbonate with dimensions of $50\times50\times50$ cm$^3$.
The total evaporative flux of the slab of hydrogel is measured from the time evolution of its weight.
A precision scale (Ohaus Pioneer 210~g) with a precision of $0.1$ mg is used with the hook available below the apparatus.
To hang the hydrogel on this hook, we made a clip with two polycarbonate rectangular plates covered with sand papers.
These two parts are placed on the top of the hydrogel and tightened with nylon screws.
This clip is then attached to the hook on the scale with a thin wire.

The scale is interfaced with a Python code using the \textit{pyserial} library to record the weight every second.
The humidity is regulated with a PID controller based on an Arduino Uno and a humidity sensor (Honeywell HIH-4021-003) positioned far from the evaporating surface.
Dry air is produced by circulating ambient air with an air pump (Tetra APS 300) in a container filled with desiccant made of anhydrous calcium sulfate (Drierite).
Moist air is obtained by bubbling air in water.
The relative humidity is set to $R_H=50$\% in all of our experiments.

\subsection{Parameters and order of magnitudes}

Buoyancy is the driving force of the free convection flow.
It is therefore necessary to estimate the density contrast between the ambient air and the air saturated in water vapor.
The air density $\rho$ at a pressure $P_0$, a temperature $T$ and a relative humidity $R_H$ is given by \cite{Tsilingiris2008}
\begin{equation}
    \rho = \frac{1}{z_m} \frac{P_0}{{\cal R} T} M_d \left[1 - f R_H \left( 1 - \frac{M_w}{M_d}  \right)  \frac{P_s}{P_0} \right],
\end{equation}
where $z_m$ and $f$ are the compressibility and enhancement factors, ${\cal R}$ the ideal gas coefficient, $M_d$ and $M_w$ the molar density of dry and saturated air, and $P_s$ the saturated pressure.
At room temperature, we have $f = 1$, $z_m = 1$ \cite{Tsilingiris2008} and a saturated pressure $P_s = 2.3$~kPa \cite{Tennent1971}.
Thus, the density contrast at a relative humidity $R_H=0.5$ is $|\rho_s - \rho_\infty| / \rho_\infty = 5 \times 10^{-3}$.
In addition, the kinematic viscosity is $\nu = 1.5\times 10^{-5}$ m$^2$/s and the diffusion coefficient of water vapor in air is ${\cal D} = 1 \times 10^{-5}$ m$^2$/s.
To justify the Boussinesq approximation used in the following Section \ref{sec:model}, it is useful to estimate the typical relative variations of the viscosity and diffusivity.
According to Tsilingiris~\cite{Tsilingiris2008}, the relative dynamical viscosity variation is $1.2\times 10^{-2}$ and the relative diffusivity variation is $4\times 10^{-3}$, which are both weak.

The Grashof number ${\rm Gr}$ is defined as the ratio between the buoyant forces driving convection and the viscous forces damping this flow, \textit{i.e.}

 \begin{equation}\label{eq:global_grashof}
     {\rm Gr} =  \frac{g}{\nu^2} \frac{\rho_\infty - \rho_s}{\rho_\infty} {\cal L}^3,
 \end{equation}
 where ${\cal L}$ is the characteristic vertical length scale of the surface of evaporation.
This Grashof number will be derived from the equations of hydrodynamics in Section \ref{sec:self-similar}.

This number predicts the dynamics of evaporation \cite{Shahidzadeh-Bonn2006,Dunn2009,Weon2011,Kelly-Zion2011,Kelly-Zion2013,Carle2013a,Somasundaram2015,Boulogne2017a}.
For small Grashof numbers, the evaporation is diffusive such that the ambient air can be considered as quiescent.
However, for large Grashof numbers, convection takes place.
The transition can be written as a critical length scale ${\cal L}^\star$ defined for ${\rm Gr}=1$, \textit{i.e.}

\begin{equation}
{\cal L}^\star = \left(\frac{g}{\nu^2} \frac{\rho_\infty - \rho_s}{\rho_\infty} \right)^{-1/3}.
\end{equation}
For water in room conditions with a relative humidity of ${\cal R}_H=50$\%, ${\cal L}^\star$ is typically $2$~mm.
Thus, for vertical film of few centimeters or more in height, convection is expected to produce an effect on the evaporative flux.

In the next Section, we recall the equations describing the natural convective flow near a vertical surface.
From these equations, we will derive the evaporative flux that can be compared with experiments.

\section{Theory}\label{sec:model}

\begin{figure}
    \centering
    \includegraphics[scale=1]{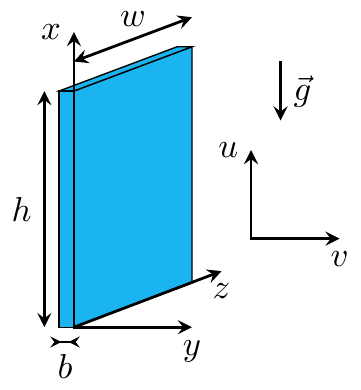}
    \caption{Sketch presenting the notations used in the theory to describe the gas flow near the vertical hydrogel represented in blue.}
    \label{fig:notations}
\end{figure}

We consider the convective evaporation of a vertical rectangular surface of width $w$, height $h$ and thickness $b$.
The model focuses on the flow due to the largest surfaces of area $wh$, referred thereafter as plates.
As a first approximation, the flow is assumed bidimensional, independent on $z$.
The effect of the two vertical edges of area $hb$ will be discussed afterwards.
The velocity field in the gas phase is denoted ${\bm u} = u {\bm e}_x + v {\bm e}_y$.
All the notations are presented in Fig.~\ref{fig:notations}.

Equations of mass transfer due to evaporation are analogous to heat transfer.
Therefore, we closely follow the derivations of Schmidt and Beckmann on heat transfer published in 1930 \cite{Schmidt1930}, also detailed later by Ostrach \cite{Ostrach1953}.
The reader may also refer to textbooks such as \cite{Bejan1993}.
We choose to rewrite in details these derivations with notations adapted to mass transfer for the convenience of the reader.

\subsection{Formulation of the problem}

We denote $\rho_s$ the density of saturated air corresponding to a vapor concentration $c_s$.
We assume that the gas phase obeys the ideal gas law so that the density $\rho$ varies linearly with the vapor concentration $c$,

\begin{equation}
    \rho(c) = \rho_0 - \Delta \rho \frac{c}{c_s},
\end{equation}
with $\Delta \rho = \rho_0 - \rho_s$ and $\rho_0$ the density of dry air.

In Cartesian coordinates, the continuity equation is

\begin{equation}\label{eq:continuity}
    \frac{\partial u}{\partial x} + \frac{\partial v}{\partial y} = 0.
\end{equation}
The momentum equation is written within the Boussinesq approximation, which considers that the gas density is constant and equal to the density $\rho_\infty$ far from the surface, except in the term driving the convection by a buoyancy effect, namely $\rho(c) g$.
Assuming a stationary flow, we have

\begin{equation}\label{eq:NS1}
    \rho_\infty \left(u \frac{\partial u}{\partial x} + v \frac{\partial u}{\partial y} \right) = - \rho(c)g + \mu \left( \frac{\partial^2 u}{\partial x^2} + \frac{\partial^2 u}{\partial y^2}\right),
\end{equation}
where $\mu$ is the gas viscosity.
The term $\partial^2 u/\partial x^2$ is negligible compared to $\partial^2 u/\partial y^2$ due to the boundary layer geometry present in the limit of large Grashof numbers.

The volatile solvent concentration field follows a diffusion-convection equation

\begin{equation}\label{eq:diff-conv1}
    u \frac{\partial c}{\partial x} + v \frac{\partial c}{\partial y} = {\cal D} \frac{\partial^2 c}{\partial y^2},
\end{equation}
where ${\cal D}$ is the diffusion coefficient of the vapor in the gas phase.

Equations (\ref{eq:continuity}-\ref{eq:diff-conv1}) must satisfy five boundary conditions.
Two are related to the concentration fields with a saturated vapor concentration near the surface and a concentration $c_\infty$ far from it.
In addition, three hydrodynamical conditions are set, with a no-slip condition on the surface for both components $u$ and $v$, and a vanishing vertical component $u$ far from the surface.
All these conditions write
\begin{equation}\label{eq:BC1}
    \begin{split}
        c(x,y=0)&=c_s,\\
        c(x,y\rightarrow\infty)&=c_\infty,
    \end{split}\qquad
    \begin{split}
        u(y=0)&=0,\\
        v(y=0)&=0,\\
        u(y\rightarrow\infty)&=0.
    \end{split}
\end{equation}

\subsection{Non-dimensionalized equations}\label{sec:self-similar}

To solve the set of equations (\ref{eq:continuity}, \ref{eq:NS1}, \ref{eq:diff-conv1}) with the boundary conditions \eqref{eq:BC1}, we non-dimensionalize the system.
First, we define the dimensionless concentration $\tilde c = (c - c_\infty)/(c_s - c_\infty)$ and the streamfunction $\Psi$ as
\begin{equation}
    u = \frac{\partial \Psi}{\partial y},\qquad  v = - \frac{\partial \Psi}{\partial x}.
\end{equation}
Therefore, equation~\eqref{eq:diff-conv1} becomes
\begin{equation}\label{eq:diff-conv2}
    \frac{\partial \Psi}{\partial y} \frac{\partial \tilde c}{\partial x} + \frac{\partial \Psi}{\partial x} \frac{\partial \tilde c}{\partial y} = {\cal D} \frac{\partial^2 \tilde c}{\partial y^2},
\end{equation}
and the momentum equation \eqref{eq:NS1} is rewritten as
\begin{equation}\label{eq:NS2}
    \frac{\partial \Psi}{\partial y} \frac{\partial^2 \Psi}{\partial x \partial y}  -
    \frac{\partial \Psi}{\partial x} \frac{\partial^2 \Psi}{\partial  y^2}  =  - \frac{\rho_\infty - \rho_s}{\rho_\infty} g \tilde c +  \nu \frac{\partial^3 \Psi}{\partial y^3},
\end{equation}
where $\nu = \mu/\rho_\infty$ and with the boundary conditions

\begin{equation}\label{eq:BC2}
    \begin{split}
        \tilde c(x,y=0)&=1,\\
        \tilde c(x,y\rightarrow\infty)&=0,
    \end{split}\qquad
    \begin{split}
        \frac{\partial \Psi}{\partial y}(x, y=0)&=0,\\
        \frac{\partial \Psi}{\partial x}(x, y=0)&=0,\\
        \frac{\partial \Psi}{\partial y}(x, y\rightarrow\infty)&=0.
    \end{split}
\end{equation}
We introduce the self-similar variable
\begin{equation}\label{eq:xi}
    \xi = \left( \frac{g}{4\nu^2} \frac{\rho_\infty - \rho_s}{\rho_\infty} \right)^{1/4} \frac{y}{x^{1/4}},
\end{equation}
that we can also write $\xi = 4^{-1/4} {\rm Gr}_x^{1/4} y/x$
 with the local Grashof number defined as
 \begin{equation}\label{eq:local_grashof}
     {\rm Gr}_x =  \frac{g}{\nu^2} \frac{\rho_\infty - \rho_s}{\rho_\infty} x^3.
 \end{equation}
For $x= h ={\cal L}^\star$, we recover the definition of the global Grashof number, as defined by Eq.~\eqref{eq:global_grashof}.

We rewrite the concentration field $\tilde c(x,y) = \hat c(\xi)$ and the streamfunction $\Psi = 4^{3/4} \nu\, {\rm Gr}_x^{1/4} \,f(\xi)$ where $f$ is the dimensionless streamfunction.
Equations \eqref{eq:diff-conv2} and \eqref{eq:NS2} become respectively

\begin{subequations}\label{eq:system_to_solve}
    \begin{align}
        \hat c'' + 3 \,{\rm Sc} \, f \hat c' = 0,\\
        f''' + 3 f f'' - 2 f'^2 + \hat c = 0,
    \end{align}
\end{subequations}
where we introduced the Schmidt number ${\rm Sc}= \nu/{\cal D}$.
The related boundary conditions \eqref{eq:BC2} become

\begin{equation}\label{eq:BC3}
    \begin{split}
        \hat c(\xi=0)=1,\\
        \hat c(\xi\rightarrow\infty)=0,
    \end{split}\qquad
    \begin{split}
        f'(\xi=0)&=0,\\
        f(\xi=0)&=0,\\
        f'(\xi\rightarrow\infty)&=0.
    \end{split}
\end{equation}

\subsection{Resolution}

\begin{figure}
    \centering
    \includegraphics[width=\linewidth]{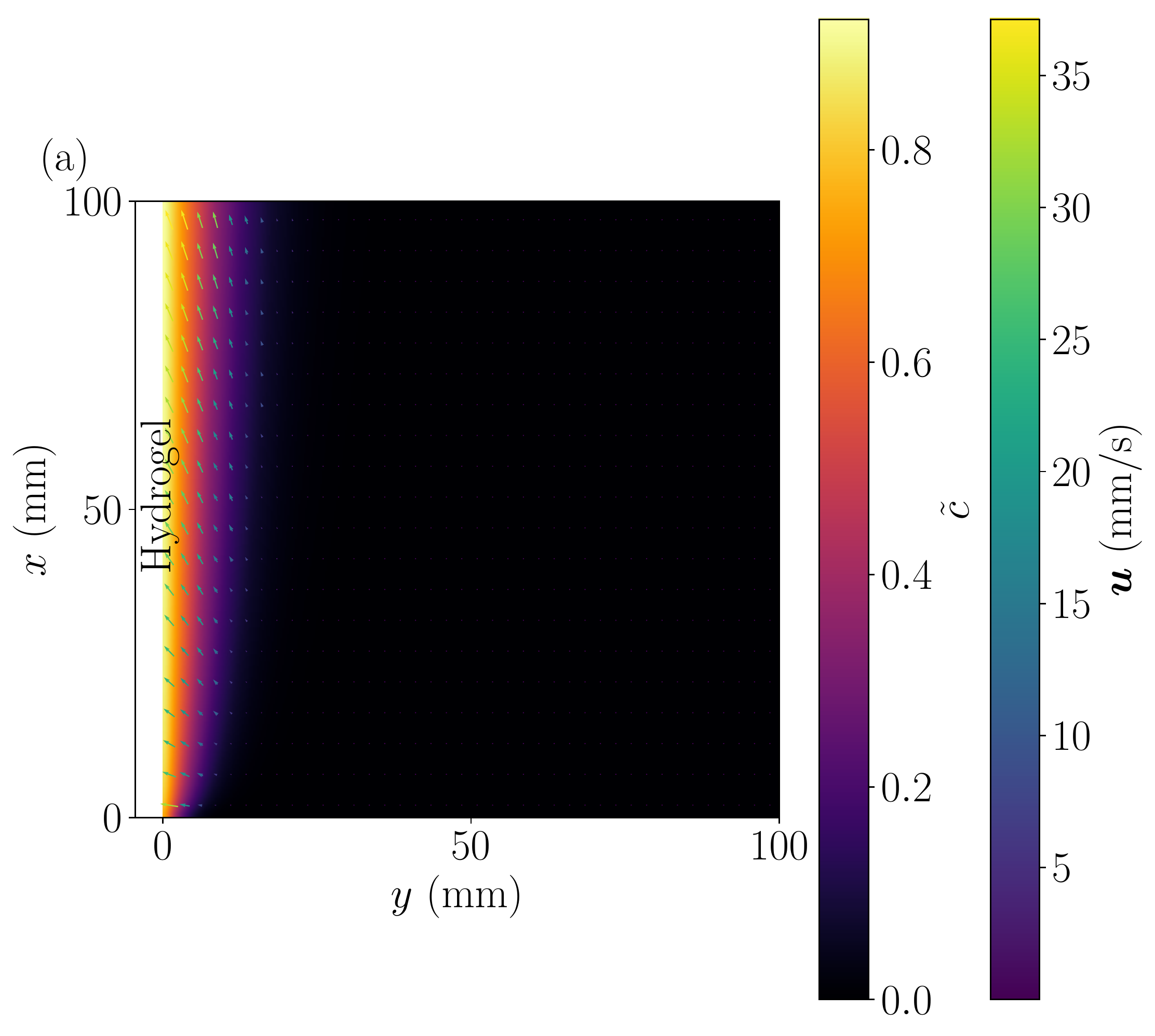}\\
    \includegraphics[width=\linewidth]{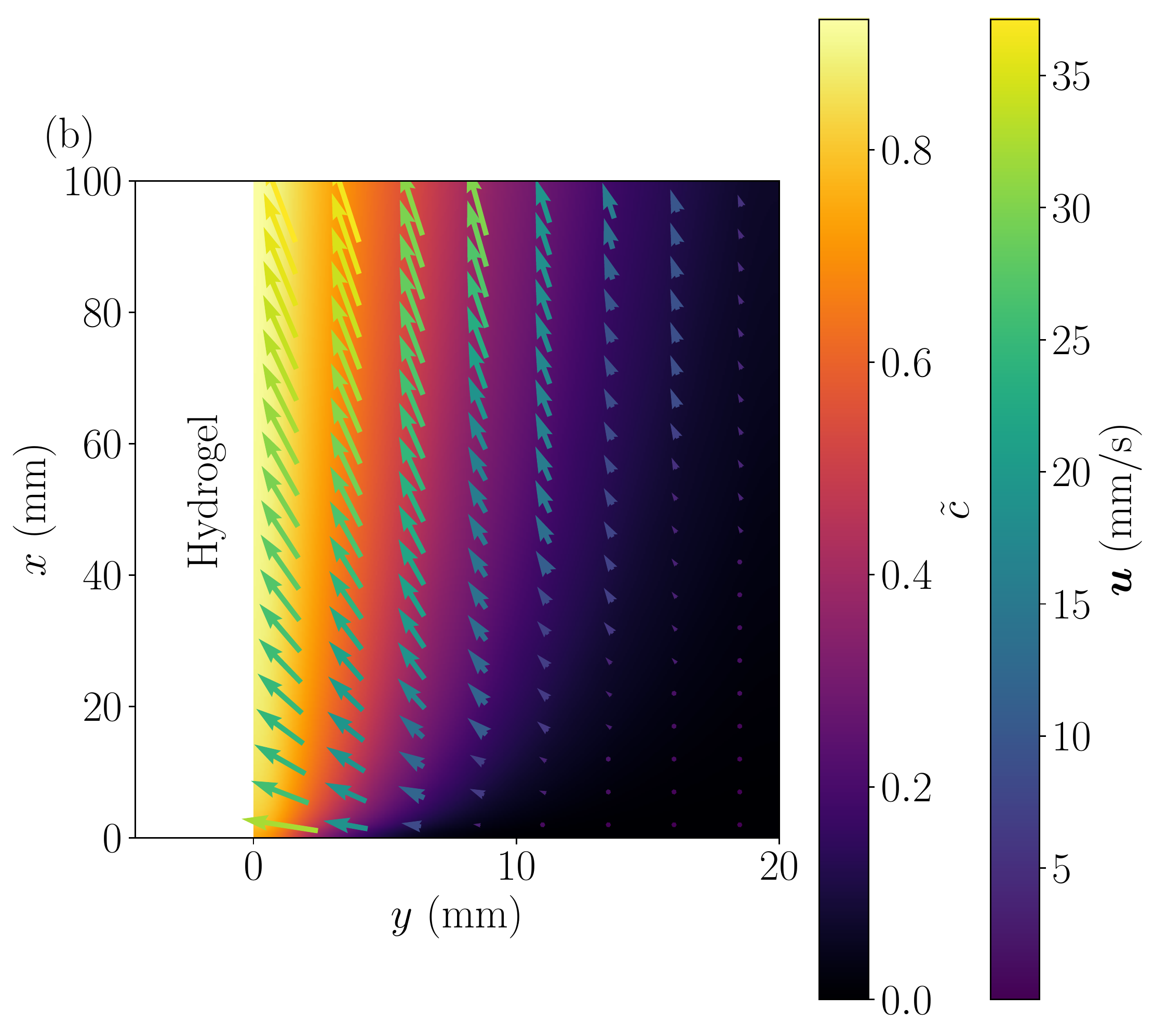}
    \caption{Concentration and velocity fields near a vertical plate with parameters corresponding to our experimental conditions obtained from a numerical integration.
    The dimensionless concentration $\tilde c$ is indicated with the colored background and the velocity field ${\bm u}$ is represented by the arrows.
    (a) The axes respect the orthonormality and (b) the $y$-axis is stretched for the sake of legibility.
    The white domain represents the hydrogel.
    }
    \label{fig:numerics}
\end{figure}

We solve the system of ordinary differential equations \eqref{eq:system_to_solve} with the conditions \eqref{eq:BC3}.
As two of these conditions are specified at $\xi\rightarrow \infty$, we use a shooting method by guessing the values of $\hat c_0' = \hat c'(\xi=0) = ({\rm d} \hat c/{\rm d} \xi)|_{\xi=0}$ and
 $\hat f''(\xi=0)$.
 A fast convergence toward a solution satisfying $\hat c(\xi\rightarrow\infty)=0$ and $f'(\xi\rightarrow\infty)=0$ is obtained with the method described by Nachtsheim and Swigert \cite{Nachtsheim1965}.
Briefly, this method consists in writing and solving differential equations on the shooting parameters to determine the corrections on the starting parameters.
We implemented this algorithm with the Python language, the libraries numpy and scipy \cite{Walt2011,Jones2001}.
Our implementation has been successfully validated with the data available in \cite{Nachtsheim1965} and our source code is available in Supplementary Materials.

The resolution is performed for ${\rm Sc} = 0.75$ and we represent the dimensionless vapor concentration $\tilde c$ and the velocity field ${\bm u}$ in Fig.~\ref{fig:numerics}.
The thickness of the boundary layer along the plate grows as $x^{1/4}$ as indicated by the self-similar variable $\xi$ defined in Eq.~\eqref{eq:xi} and the flow is oriented mainly upward with a horizontal component toward the plate.
In addition, we obtain $\hat c_0'\approx -0.51$ for our conditions.
The value of this coefficient is used hereafter in Sec.~\ref{Sec:evaporative_flux} to estimate the evaporative flux.

\subsection{Evaporative flux} \label{Sec:evaporative_flux}

As the concentration field surrounding the plate is now determined, we can calculate the evaporative flux.
The local flux is defined as
\begin{equation}
    j = - {\cal D} \left.\frac{\partial c}{\partial y}\right|_{y=0}.
\end{equation}
With the definition of the dimensionless concentration $\hat c$, we obtain, on the surface of the plate of area $wh$, the local convective flux

\begin{equation}\label{eq:local_convective_flux}
    j_{\rm plate} = - {\cal D} (c_s - c_\infty) \hat c_0' \frac{{\rm Gr}_x^{1/4}}{x}.
\end{equation}
Note that the numerics is only necessary to determine the prefactor $\hat c_0'$ of Eq.~\eqref{eq:local_convective_flux} and that the scaling is obtained analytically.

The total evaporative flux due to convection is $Q_{\rm plate} = \int j_{\rm plate} \,{\rm d} S$, which leads after integration to

\begin{equation}\label{eq:Q_conv}
    Q_{\rm plate} = 2 w {\cal D} (c_s - c_\infty) \left(- \frac{4}{3} \hat c_0'\right) {\rm Gr}^{1/4},
\end{equation}
where ${\rm Gr}$ is defined by equation \eqref{eq:global_grashof} for the characteristic length scale ${\cal L} = h$.

In the next Section, we discuss the validity of this prediction that we compare to our experimental results obtained by drying vertical hydrogels of different sizes and aspect ratios.

\section{Discussion}

\begin{figure}
    \centering
    \includegraphics[scale=1]{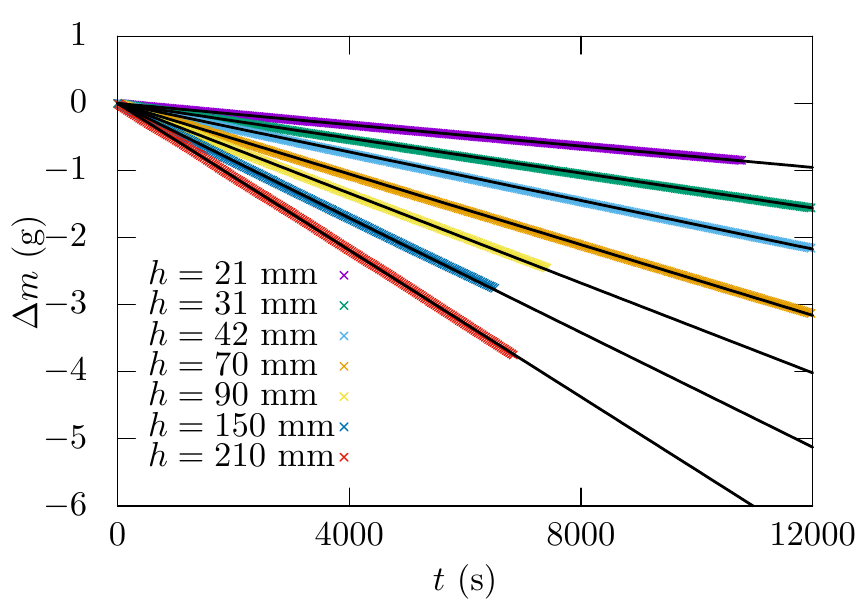}
    \caption{Raw data of the time evolution of the relative weight $\Delta m$ for $w=80$~mm and for different hydrogel height $h$.
    Black solid lines are linear fits of the different sets of data.}
    \label{fig:raw_data}
\end{figure}

In Fig.~\ref{fig:raw_data}, we present data on the time variation of the relative weight $\Delta m$ for some of the hydrogels used in our experiments.
To obtain the total flux $Q$, we fit our measurements with a linear law $\Delta m(t) = -Qt$.
This procedure is repeated for hydrogels of different sizes.
The linear fits indicate that the evaporative flux remains constant over the duration of the experiment.
Thus, the variations of the polymer concentration in the hydrogel is negligible enough to ensure a constant saturated vapor concentration.
In a previous work, we also checked that this saturated vapor concentration can be assimilated to the one of pure water \cite{Boulogne2017a}.

\begin{figure}
    \centering
    \includegraphics[scale=0.95]{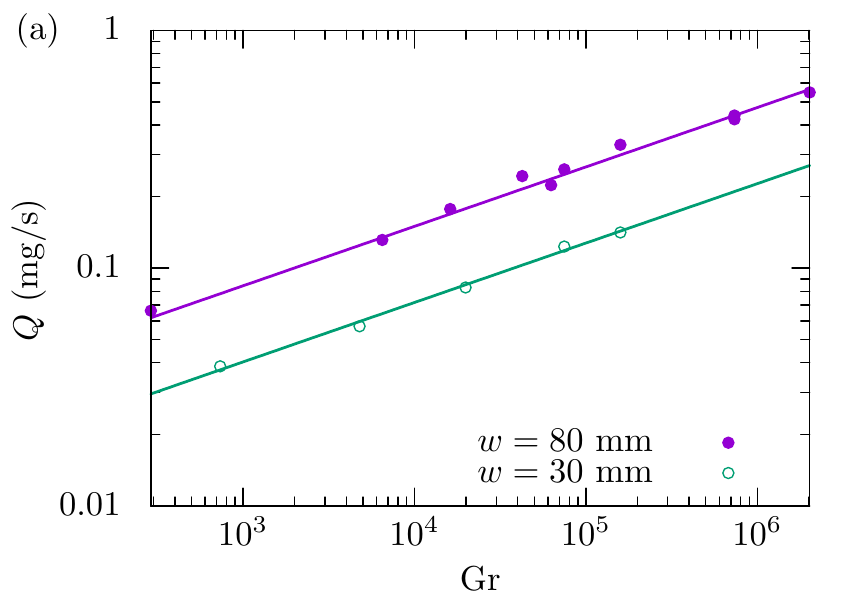}\\
    \includegraphics[scale=0.95]{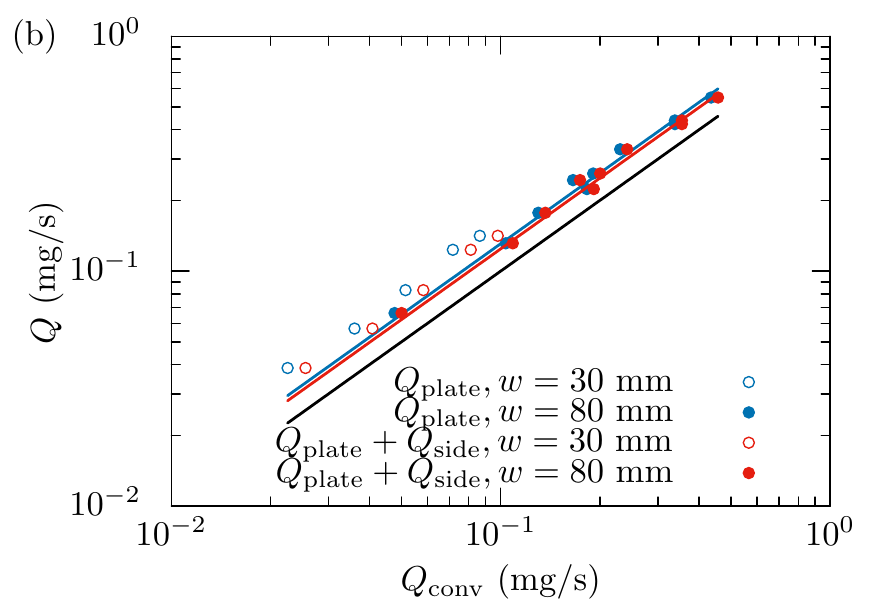}
    \caption{(a) Experimental evaporative flux $Q$ as a function of the Grashof number ${\rm Gr}$ for two gel widths $w$.
Solid lines are fits in ${\rm Gr}^{1/4}$.
    (b) Comparison between the experimental fluxes $Q$ and the theoretical prediction $Q_{\rm conv}$ from Eqs~\eqref{eq:Q_conv} and Eq \eqref{eq:Q_conv_with_b}.
    Open symbols are for $w=30$~mm and closed symbols for $w=80$~mm.
    Blue points correspond to $Q_{\rm conv}=Q_{\rm plate}$ and red points to $Q_{\rm conv}=Q_{\rm plate} + Q_{\rm side}$.
    Each colored solid line is a linear fit on all data points of the same color and the black solid line represents $Q = Q_{\rm conv}$.}
    \label{fig:compare_ex_theory}
\end{figure}

In Fig.~\ref{fig:compare_ex_theory}(a), we show experimental measurements of the evaporative flux $Q$ for two hydrogel widths, $w=30$~mm and $w=80$~mm and heights ranging between $15$~mm and $210$~mm.
We plot these fluxes as a function of the Grashof number ${\rm Gr}$, which is varied over three orders of magnitude.
As expected, the evaporative flux is systematically larger for wider gels at a given Grashof number, \textit{i.e.} at a given height $h$.
For both widths, the measurements show a good agreement with the prediction in ${\rm Gr}^{1/4}$ obtained from Eq.~\eqref{eq:Q_conv}, which means that the power law on the Grashof number is well captured by the model.

In Fig.~\ref{fig:compare_ex_theory}(b), we compare our measurements with the prediction given by Eq.~\eqref{eq:Q_conv}.
By fitting all our measurements $Q$ to the prediction $Q_{\rm conv}\simeq Q_{\rm plate}$, we obtain $Q\approx 1.31 \,Q_{\rm conv}$, which is particularly reasonable at this level of theoretical description and for the precision of our experiments.
Nevertheless, in this representation, we can notice for the smallest hydrogel widths that $Q_{\rm conv}$ underestimate the measured flux.
We interpret this deviation by the absence of consideration of edge effects in the theoretical description that become visible for our smallest samples.
Edge effects may arise from the two side edges and the bottom edge.
Also, tip effects at the corners introduce a divergence in the local flux, which is not taken into account.
The major difficulty in describing all these effects is that they are not simply additional corrections because edges and tips can have a mutual influence.

Despite these difficulties, we attempt to write some corrections to the prediction given by Eq.~\eqref{eq:Q_conv}.
First, we consider that the local flux $j_{\rm conv}$ obtained by considering the surfaces of area $wh$, also holds for the two vertical surfaces of area $bh$.
Thus, the integral is calculated over a total area $2(w + b)h$ and the flux due to convection $Q_{\rm conv} \simeq Q_{\rm plate} + Q_{\rm side}$ is
\begin{equation}\label{eq:Q_conv_with_b}
      Q_{\rm conv} \simeq 2 (w+b) {\cal D} (c_s - c_\infty) \left(- \frac{4}{3} \hat c_0'\right) {\rm Gr}^{1/4}.
\end{equation}
In Fig.~\ref{fig:compare_ex_theory}(b), we observe that this correction reduces the deviation for $w=30$~mm and provide an estimate closer to the experimental results, $Q\approx 1.24 \,Q_{\rm conv}$.

A second correction that could be considered is related to the bottom edge.
In our previous study \cite{Dollet2017}, we argued that the boundary layer approximation breaks down at the leading edge of the plate \cite{Leal2007} over a length scale ${\cal L}^\star$.
This correction must be also related to the evaporative flux over the bottom side of area $bw$.
Applying these small corrections lead to a better agreement with the measurements but, presently, their validities are particularly difficult to discuss.
Thus, we prefer to do not overinterpret our results as the main trend is well captured.
Further developments, especially from numerical simulations, could help to refine the model on these aspects.

\section{Conclusion}

Motivated by the question of the soap film bursting, we identified that one of the key aspects is related to the evaporation of the soap film.
As shown by the recent literature, evaporation can induce a convective flow in the atmosphere, which depends on the geometry and influences the evaporation kinetics.
Therefore, to understand the effect of evaporation of soap films, it is crucial to develop such models.

In this paper, we followed derivations established for heat transfer to calculate the evaporative flux on a vertical plate.
To verify the validity of this description, we performed model experiments with slabs of hydrogels of various widths and heights.
From the time evolution of the weight, we measured the total evaporation rate that we compared to the theoretical prediction.
We show that our experiments are in good agreement with this model, following the power law ${\rm Gr}^{1/4}$.
Although that the model does not consider edge and tip effects, it provides a reasonable estimation of the prefactor.
We indicate that corrections could lead to a better agreement, especially for small films where edge effects are more significant.
We suggest that a more complete description including these effects would require numerical simulations.

Nevertheless, our results shows that the model presented in this paper provides a good description of the evaporation of vertical rectangular films, provided that the boundary layer remains laminar all along the surface.
In the near future, we would like to investigate the evaporation of soap films in light with this theory to rationalize the mechanisms that relates the bursting of soap films to their evaporation.
We expect that different additional mechanisms can have a contribution to the evaporation of soap films, which are not present in the model system based on hydrogels.
First, the composition of the soap solution can lead to a vapor pressure different from water, eventually with an effect of the increasing concentration of non-volatile molecules. 
The evaporation of thin films may generate temperature gradients, and therefore thermal Marangoni flows in the soap film, in addition to the solutal Marangoni flows, the gravitational drainage, and the marginal regeneration in free standing films.
The film thinning due to evaporation and liquid flow leads the appearance of the so-called black film where van der Waals attraction forces balance the double-layer repulsion forces, and therefore modifies the chemical potential of the liquid \cite{Belorgey1991}.

\section*{Acknowledgments}
We are particularly grateful to Emmanuelle Rio for the initial discussions that motivated this study.
We also thank Lor\`ene Champougny, Jonas Miguet, Christophe Poulard and Fr\'ed\'eric Restagno for discussions.
F.B. thanks M\'elanie Decraene for her assistance.
This study has been carried out with a funding support by the ANR (ANR-11-BS04-0030-WAFPI project).

\bibliography{biblio}

\bibliographystyle{unsrt}

\end{document}